\documentclass[longbibliography,prl,aps,twocolumn,showpacs,floatfix]{revtex4-2}
\usepackage{amsfonts}
\usepackage{amssymb}
\usepackage{hyperref}
\usepackage{graphicx}
\usepackage{dcolumn}
\usepackage{bm,amsmath,verbatim}
\usepackage{mathrsfs}
\usepackage{color}
\usepackage{lineno}
\usepackage{dsfont}
\usepackage{amsmath,amssymb,amsthm}

\usepackage[T1]{fontenc}
\usepackage[latin9]{inputenc}
\usepackage{booktabs}

\hypersetup{colorlinks,
	linkcolor=blue,          citecolor=blue,        filecolor=blue,      urlcolor=blue           }

\def\be{\begin{equation}}
	\def\ee{\end{equation}}
\def\bea{\begin{eqnarray}}
	\def\eea{\end{eqnarray}}
\def\bse{\begin{subequations}}
	\def\ese{\end{subequations}}

\def\be{\begin{eqnarray}}
	\def\ee{\end{eqnarray}}

\begin{document}
\title{Nonlinearity-Induced Thouless Pumping in Quasiperiodic Lattices}
\author{Xiao-Xiao Hu$^{1,4}$}
\email[]{Contact author: dxhuxiaoxiao@163.com}
\author{Dun Zhao$^{2}$}
\author{Hong-Gang Luo$^{3,4}$}
\email[]{Corresponding author: luohg@lzu.edu.cn}
\affiliation{$^{1}$School of Physical Science and Technology, Lanzhou University, Lanzhou 730000, China}
\affiliation{$^{2}$School of Mathematics and Statistics, Lanzhou University, Lanzhou 730000, China}
\affiliation{$^{3}$Institute of Fundamental Physics and Quantum Technology, Ningbo University, Ningbo, 315211 China}
\date{\today}
\affiliation{$^{4}$Lanzhou Center for Theoretical Physics $\&$ Key Laboratory of Theoretical Physics of Gansu Province, Lanzhou University, Lanzhou 730000, China}
\begin{abstract}
		 Nonlinear Thouless pumping has been established in periodic lattices; its counterpart in quasiperiodic lattices remains unexplored. Here, we show a nonlinear topological pumping of gap solitons in quasiperiodic lattices where the local nonlinear self-consistent potentials lead to a lattice potential reconstruction; as a result, an emergent topological structure induced by this local reconstruction governs the dynamics of the gap solitons. This enables solitons to adiabatically occupy a single topological band, realizing quasi-quantized Thouless pumping. In addition, the intrinsic lattice perturbations disrupt this band occupation, which drives solitons into a non-quantized drifting regime. However, even in this regime, we also find that the soliton transport is constrained by the topological properties of a critical rational approximant. Tuning nonlinearity or lattice scaling reveals a controllable switching among topological pumping, drifting, and localization. Our work uncovers a mechanism for nonlinearity-induced topological behavior in complex lattice potentials.  
\end{abstract}
\maketitle
Thouless pumping is a topologically protected transport mechanism~\cite{Thouless1983,Niu1984,Di2010,Wang2013,Ke2016,Marra2015,Ke2017,Lin2020}, where the quantized and robust flow of a physical observable emerges from the adiabatic cyclic evolution of a system with isolated energy bands~\cite{Lohse2016,Nakajima2016,Walter2023,Zilberberg2018,Cerjan2020,Grinberg2020,Xia2021,Ma2018,Cheng2020,Fedorova2020,Nakajima2021}---the spectral gaps between them providing the necessary protection for topological invariants~\cite{Niu1984}. This phenomenon can be understood by two fundamental pillars: the presence of translational symmetry, which enables a band description, and the linear band topology characterized by integer Chern numbers~\cite{Thouless1983,Niu1984,Di2010}. Recently, further explorations have extended topological pumping into the nonlinear regime within periodic potentials~\cite{Jurgensen2021,Jurgensen2022,Jurgensen2023,Fu2022,Fu2022t,Mostaan2022,Tuloup2023,Citro2023,Hu2024,Lyu2024,Szameit2024,Cao2024,Cao2025,Kartashov2025,Ye2025,Tao2025,Fleischhauer2026,Zhou2026}. There, solitons under adiabatic driving exhibit quantized nonlinear Thouless pumping~\cite{Jurgensen2021,Jurgensen2023}, where the integer or fractional transport of solitons is understood as the flow of the instantaneous maximally localized single-band (or multi-band) Wannier functions of the linear Hamiltonian~\cite{Jurgensen2022,Jurgensen2023,Mostaan2022}. Yet a pivotal finding has recently emerged: this quantization endures even when the linear bands governing the evolution are themselves topologically trivial~\cite{Tao2025}. This reveals that nonlinearity can sustain quantization beyond the confines of linear band topology~\cite{Tao2025,Fleischhauer2026}, prompting a fundamental inquiry: what becomes of nonlinear Thouless pumping in the absence of translational invariance? Quasiperiodic lattices, which interpolate between order and disorder, offer a unique testbed~\cite{Nakajima2021,Ye2024}. They host rich topological phases~\cite{Nakajima2021,Oded2012,Marra2020} and fractal energy spectra~\cite{Weld2015,Anuradha2021}, yet their absence of discrete translational symmetry seemingly precludes the very notion of a unit cell and, consequently, renders the existing nonlinear Thouless pumping theory in periodic potentials not directly applicable~\cite{Jurgensen2023,Tao2025}.

Here we address the nonlinear pumping of gap solitons formed in a BEC with attractive atomic interactions loaded in a slowly varying quasiperiodic lattice. We find that solitons reconstruct the local lattice potential via their density distribution, giving rise to an emergent topological structure within the localized region. This enables solitons to almost occupy a single topological band and maintain this occupancy throughout the adiabatic cycle, allowing them to follow the corresponding Wannier functions and realize nonlinear Thouless pumping. When this occupancy is disrupted by perturbations arising from period mismatches, interband transitions drive the soliton into a non-quantized drifting regime; nevertheless, the drift direction remains constrained by the topology of a critical rational approximant. By tuning the nonlinearity or lattice scale to modify the local potential induced by the soliton, one can controllably switch among topological pumping, drifting, and localization. These findings hold promise for application in platforms such as ultracold atomic gases and photonics.

\emph{Model}---We consider a one-dimensional condensate in an optical lattice, described by the dimensionless Gross-Pitaevskii (GP) equation
\begin{figure*}[tp]
	\includegraphics[width=1.01\linewidth]{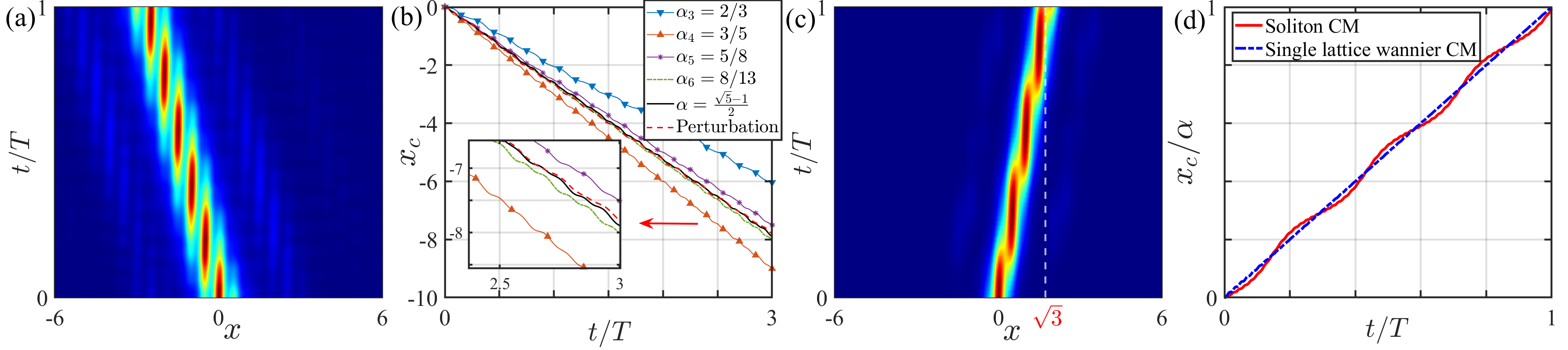}
	\centering
	\caption{Gap soliton pumping in a quasiperiodic superlattice. (a)--(b) For $\alpha=\frac{\sqrt{5}-1}{2}$: (a) Temporal evolution of the soliton density. (b) Soliton center-of-mass trajectories for successive rational approximants. These trajectories converge beyond the fifth-order approximant. The red dashed line is the trajectory for the fifth-order approximant including the perturbation $W_n^{\infty}$. (c)--(d) For $\alpha=\sqrt{3}$: (c) Temporal evolution of the soliton density. (d) Soliton center-of-mass trajectory (solid line) compared with the instantaneous Wannier center of a sliding lattice (dashed line), showing adiabatic following. Other parameters: $p_1=p_2=25$, $N=0.2$, $v=0.1$, $T=10\pi$.}
	\label{fig1}
\end{figure*}	
\begin{equation}\label{GPE}
	i \frac{\partial}{\partial t}\Psi(x,t)=H(x,\varphi)\Psi(x,t)-|\Psi(x,t)|^{2} \Psi(x,t).
\end{equation}
Here, $H=-\frac{1}{2}\partial_{xx}+V(x,\varphi)$ is the linear Hamiltonian, with the $\varphi$-dependent superlattice potential~\cite{Lohse2016,Nakajima2016}
 \begin{equation}\label{eq2}
	V(x,\varphi)=-p_{1} \cos ^{2}\left(2\pi x\right)-p_{2} \cos ^{2}\left(\frac{\pi x}{\alpha }+\varphi(t) \right),
\end{equation}
where $p_{1,2}$ are the dimensionless depths of the two constitutive lattices. The parameter $\alpha=d_2/(2d_1)$, where $d_{1,2}$ are the real-space periods of the individual lattices~\cite{Lohse2016}, governs the overall periodicity of $V(x,\varphi)$. For rational $\alpha$ ($2\alpha=p/q$ with positive coprime integers $p,q$), the superlattice potential is periodic with period $L=q\alpha=p/2$, while for irrational $\alpha$ it is quasiperiodic. The phase $\varphi(t)=-vt$ is varied slowly (small $v$) to ensure adiabaticity over an adiabatic period $T=\pi/v$. In experiments, energy, length, and time are measured in units of $E_0=\frac{\hbar^2}{4md_1^2}$, $x_0=2d_1$, and $t_0=\frac{\hbar}{E_0}$, respectively, where $m$ is the atomic mass. For a BEC with negative $s$-wave scattering length ($a_s<0$) in a cigar-shaped trap with transverse frequency $\omega_{\perp}$, the dimensionless $\Psi$ and the dimensional $\Phi$ order parameter are related by $\Psi=\sqrt{2\hbar\omega_{\perp}|a_s|/E_0}\Phi$ (see, e.g.,~\cite{P2025}) and the dimensionless norm is given by $N=\int_{-\infty}^{\infty}|\Psi\left(x\right)|^{2}\mathrm{d}x=\frac{\hbar\omega_{\perp}|a_{s}|}{E_{0}d_1}\tilde{N}$, where $\tilde{N}$ is the actual number of atoms. 

For a static superlattice potential ($v=0$), Eq.~(\ref{GPE}) supports a rich family of gap solitons, which have been studied in both continuous and discrete systems~\cite{Eiermann2004,Zhang2009,Zhang2009T,Vicencio2009,Dovgiy2015,Andrey2002,Sakaguchi2006,Prates2022}. Using Newton's method, we compute gap soliton solutions bifurcating from the lowest band, which then serve as the initial states in our study. Due to the localized nature of the solitons and the vanishing of $\Psi$ as $|x| \to \infty$, the linear properties of the Hamiltonian $H$ become particularly important~\cite{Jurgensen2023,Fu2022}. In particular, the quantized Thouless pumping of soliton arises from the soliton's dominant occupation of a single Bloch band~\cite{Fu2022}. By expanding the soliton wave function as $\Psi(x,t) = \sqrt{N} \sum_{n,m} a_{n,m}(t) w_{n,m}(x,t)$, we obtain the band occupation $\rho_n(t) = \sum_m |a_{n,m}(t)|^2$, where $n$ is the band index, $m$ labels the unit cell, and $w_{n,m}(x,t)$ are the instantaneous Wannier functions.

\emph{Nonlinear Thouless Pumping of Gap Solitons}---When the soliton resides predominantly in a single Bloch band ($\rho_n(t) \approx 1$), a single-band approximation is valid, and the wave function is well described by $\Psi(x,t) \approx \sqrt{N} \sum_m a_m(t) w_m(x,t)$. Under this condition, the soliton's center-of-mass motion decomposes into two distinct contributions: $x_c(t) \approx X(t) + \Delta(t)$. Here, $X(t) = \int_{-\infty}^\infty x|w_{m_f}(x,t)|^2\mathrm{d}x$ is the Wannier center that the soliton follows, whose net displacement per adiabatic cycle is quantized by the Chern number of the occupied band~\cite{Nakajima2016,Fu2022}. The second term,
\begin{equation}
	\begin{aligned}
		\Delta(t)=&\sum_{m^{\prime}} |a_{m^{\prime}}(t)|^2(m^{\prime}-m_f)L \\
		&+\sum_{m,m^{\prime} \atop m\neq m'}a_{m^{\prime}}^*(t)a_m(t)\int_{-\infty}^\infty w_{m^{\prime}}^*(x,t)xw_{m}(x,t)dx ,
	\end{aligned}
\end{equation}
represents a dynamical offset arising from multi-site effects beyond the single Wannier center description. When the total potential $V_{\text{total}}(x)=V(x) + |\Psi(x)|^2$ possesses spatial inversion symmetry, $\Delta=0$~\cite{SM2025}; however, an adiabatic periodic drive $\phi(t)$ breaks this symmetry, and the resulting lattice modulation drives $\Delta(t)$ to exhibit bounded oscillations~\cite{SM2025}. 

In a periodic superlattice potential, discrete translational symmetry ensures that the dynamical offset $\Delta(t)$ is reset after an integer number of adiabatic cycles~\cite{SM2025}, so that $\Delta(T_r) = \Delta(0)$ with $T_r = q_r T$ ($q_r$ a positive integer) defining the effective pumping period~\cite{Jurgensen2023,Tao2025,SM2025}. The net soliton displacement over $T_r$ is $x_c(T_r)-x_c(0) = X(q_r T)-X(0)$, which is quantized by the Chern number, yielding quantized integer nonlinear Thouless pumping, a robust transport mechanism insensitive to disorder and impurities~\cite{Cao2024,Rechtsman2025}.

We now extend the analysis to a quasiperiodic potential with incommensurability ratio $\alpha$. Under adiabatic driving of $\phi(t)$, we track the soliton's center-of-mass motion $x_c(t)=N^{-1}\int x|\Psi(x,t)|^2dx$ and find that soliton transport persists even in the quasiperiodic setting. Unlike the nonlinear Thouless pumping observed in periodic potential, the soliton transport here exhibits two distinct behaviors: one is non-quantized drift [Fig.~\ref{fig1}(a)(b)], where the transport per adiabatic cycle is no longer quantized; the other is quasi-quantized pumping, with a mean displacement per cycle close to $\alpha$ [Fig.~\ref{fig1}(c)(d)].

\emph{Non-quantized Drift of Gap Solitons}---To analyze the non-quantized drift in Figs.~\ref{fig1}(a) and (b), we employ periodic approximations based on the continued fraction expansion $\alpha=[a_0;a_1,a_2,...]=a_0+1/[a_1+1/(a_2+...)]$. Truncation yields rational approximants $\alpha_n = [a_0; a_1, \dots, a_n]$, converging to $\alpha$ as $n$ increases~\cite{Ye2024,Prates2022}. For high-order approximants with $|\alpha_{m}-\alpha_n|\ll1$ ($m>n$), we expand the potential difference $V_m-V_n$ and express the $m$-th approximant Hamiltonian as $H_{m}=H_n+W_n^m$, where
\begin{equation}\label{eq3}
	W_n^m=p_2\sin(2k_nx-2\varphi(t))\delta k_n^{m}x+\mathcal{O}(\delta k_n^{m}),
\end{equation}
with $k_n=\frac{\pi }{\alpha_n}$ and $\delta k_n^{m}= k_m-k_n$. The term $W_n^m$ acts as a long-wavelength tilting perturbation that modifies the periodicity of $V_n$. In the limit $m \to \infty$, the true quasiperiodic system is described by $H = H_n + W_n^{\infty}$, with $\delta k_n^{\infty}=\pi/\alpha - k_n$. Figure~\ref{fig1}(b) reveals a critical order $n_c$ in the rational approximation, separating two dynamical regimes: integer-quantized nonlinear Thouless pumping for $n \leq n_c$ (e.g., $\alpha_3, \alpha_4, \alpha_5$), and a non-quantized drift for $n>n_c$, where the soliton's center-of-mass displacement converges to a common value. This convergence signals the breakdown of quantization in the quasiperiodic limit. 

\begin{figure}[tp]
	\includegraphics[width=1.05\linewidth]{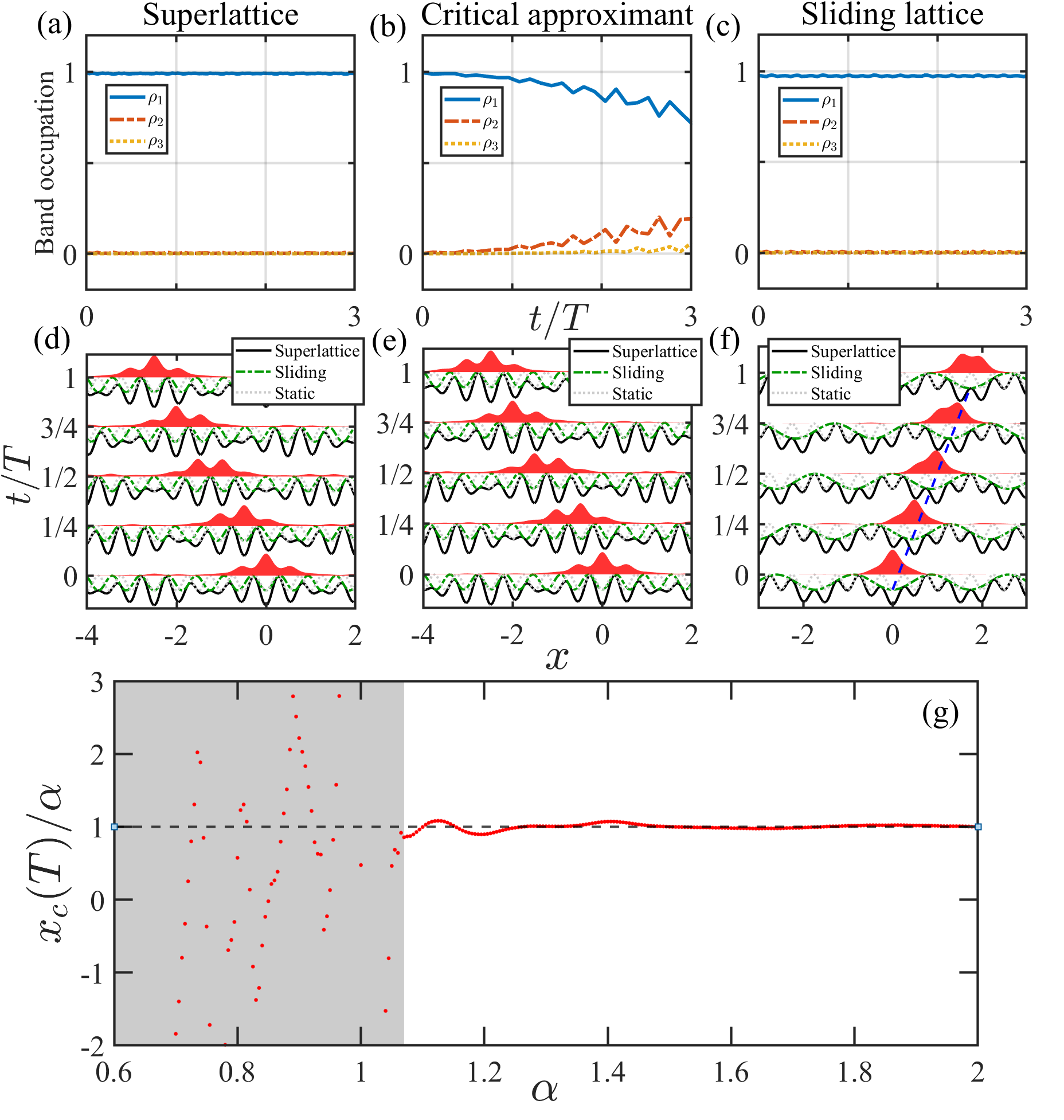}
	\caption{(a)--(c) Band occupation case of the adiabatically pumped gap soliton, expanded on the instantaneous Wannier basis derived from: (a) a linear superlattice (with $\alpha=5/8$), (b) a rational-approximation lattice at the critical order $\alpha_5$ (where $\alpha=\frac{\sqrt{5}-1}{2}$), and (c) a sliding single lattice (with incommensurate parameter $\alpha=\sqrt{3}$). (d)--(f) The shapes of the propagated solitons at $t=0$ and after each quarter adiabatic period ($T/4$) for (d) $\alpha=5/8$, (e) $\alpha=\frac{\sqrt{5}-1}{2}$ and (f) $\alpha=\sqrt{3}$. The lines at the bottom represent the lattice potential at $t=0$ and after each quarter adiabatic period. (g) The relationship between the center displacement of soliton in an adiabatic cycle and the parameter $\alpha$. Other parameters: $p_1=p_2=25$, $N=0.2$, $v=0.1$, $T=10\pi$. } 
	\label{fig2}
\end{figure}

In the quantized regime ($n \leq n_c$), the soliton initially almost fully occupies the lowest band and remains within it throughout the adiabatic cycle [Fig.~\ref{fig2}(a)], with its center of mass adiabatically tracking the instantaneous Wannier center~\cite{Jurgensen2022,Jurgensen2023,Fu2022}. As illustrated in Fig.~\ref{fig2}(d), the soliton density distribution repeats with period $T_r = T$. Consequently, the dynamical offset resets after each adiabatic cycle, $\Delta(T) = \Delta(0)$, so that the net soliton displacement equals that of the Wannier center, yielding quantized transport.

In contrast, within the quasiperiodic limit (or its high-order approximants $V_m$, $m>n_c$), the initial soliton is strongly localized on a scale much smaller than the potential period [Fig.~\ref{fig2}(e)]. At this scale, as shown in Fig.~\ref{fig2}(b), the initial soliton state projects strongly onto the lowest band of the critical Hamiltonian $H_{n_c}$ (where $H_m=H_{n_c} + W_{n_c}^m$). However, as illustrated in Fig.~\ref{fig1}(b), the resulting soliton's center-of-mass displacement is not quantize; instead, it converges to a value distinct from the quantized pumping observed in $H_{n_c}$ alone. The breakdown of quantization originates from the perturbation $W_{n_c}^m$. Acting as a long-wavelength tilting potential, it induces interband transitions, disrupts the effective periodicity of $H_{n_c}$, and causes the soliton to gradually lose its adiabatic connection to the topological band. Over the pumping cycle, the soliton progressively loses its occupancy of the lowest band [Fig.~\ref{fig2}(b)] and, driven by the tilted potential, accumulates a net spatial shift, leading to non-quantized drift [Figs.~\ref{fig2}(d)(e)]. The consistent displacement across all high-order approximants ($m > n_c$) reflects the universal role of $W_{n_c}^m$ as the dominant symmetry-breaking perturbation. This is confirmed by adding $W_{n_c}^m$ to the $H_{n_c}$ system (e.g., introducing $W_5^\infty$ into $H_5$), which successfully reproduces the non-quantized drift observed in the quasiperiodic case [Fig.~\ref{fig1}(b)]. Although the quantization of displacement is destroyed, the net displacement direction remains governed by the topology of the critical band~\cite{SM2025}. 
 
\emph{Quasi-quantized and Fractional Thouless Pumping}---In stark contrast to the non-quantized drift in quasiperiodic systems and the quantization enabled by translational symmetry in periodic potentials, increasing $\alpha$ establishes an adiabatic connection between the soliton and the lowest eigenband of the sliding lattice, giving rise to quasi-quantized pumping [Figs.~\ref{fig1}(c)(d)]. The soliton predominantly occupies this lowest band ($\rho_1 \approx 1$) [Fig.~\ref{fig2}(c)], and its center-of-mass closely follows the corresponding Wannier center [Fig.~\ref{fig1}(d)]. However, the quasiperiodic potential lacks translational invariance, so the system cannot return to an equivalent initial state after each adiabatic cycle [Fig.~\ref{fig2}(f)]; consequently, $\Delta(t)$ does not effectively reset after each cycle~\cite{SM2025}, exhibiting quasi-quantized behavior.
  \begin{figure*}[tp]
	\includegraphics[width=1.03\linewidth]{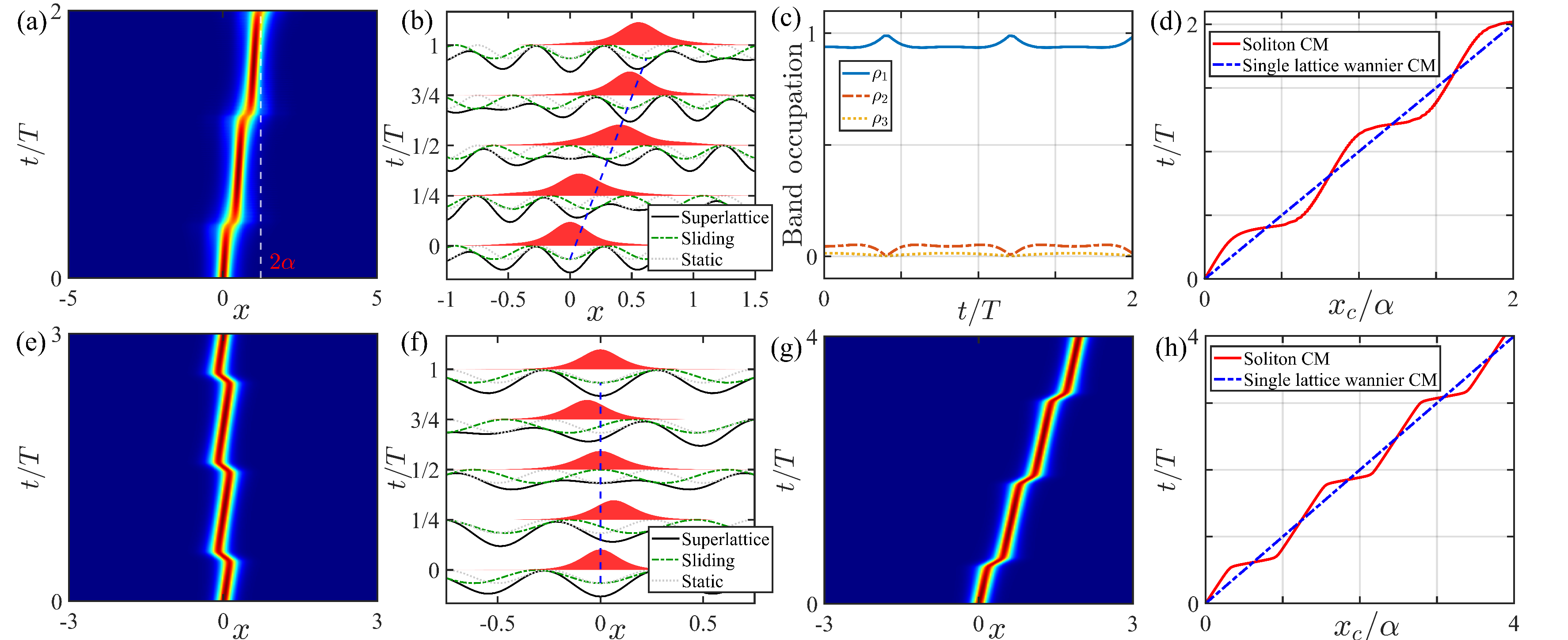}
	\caption{Nonlinearity-induced quasi-quantized pumping and trapping in a quasiperiodic superlattice. Upper panels ($N=7$): (a), (b) Temporal evolution of the soliton density and profile under adiabatic driving, showing directed transport. (c) Projection weights of the soliton wave function onto each band in the maximally localized Wannier basis of the sliding lattice. (d) Soliton center-of-mass trajectory (solid line) and the Wannier center trajectory (dashed line). Lower panels ($N=20$): (e), (f) Temporal evolution of the soliton density and profile under adiabatic driving, showing trapping. (g), (h) With a fixed long-period lattice and a sliding short-period lattice, the soliton still exhibits quasi-quantized pumping. Comparing the upper and lower panels shows that the nonlinear strength controls the transition between quasi-quantized pumping and localization. Other parameters: $p_1=p_2=15$, $\alpha=\frac{\sqrt{5}-1}{2}$, $v=0.1$, $T=10\pi$.} 
	\label{fig3}
\end{figure*}	

Further analysis reveals that although the soliton closely follows the Wannier center of a single sliding lattice, its quasi-quantized pumping is not mere mechanical translation. The essence lies in the fact that the nonlinear soliton actively modifies the local potential via its self-consistent potential, effectively reconstructing the underlying lattice. This nonlinear reconstruction mechanism enables the soliton to adiabatically follow Wannier states, thereby establishing and maintaining the adiabatic connection to the lowest band of the sliding lattice.

As shown in Fig.~\ref{fig2}(g), the displacement of the soliton's center-of-mass per adiabatic cycle as a function of $\alpha$ reveals a dynamical phase transition. In the shaded region, soliton transport corresponds to quantized nonlinear Thouless pumping in periodic systems and non-quantized drift in quasiperiodic (or large-period) potentials; by contrast, the non-shaded region exhibits nonlinear Thouless pumping induced jointly by lattice scale and nonlinearity. During each adiabatic cycle, the soliton's center-of-mass displacement is confined near the Wannier center displacement, $C\alpha$, of its predominantly occupied band (with Chern number $C=1$ for the sliding lattice's band~\cite{Lohse2016}). Unlike the quasi-quantized case, in periodic systems with rational $\alpha$, discrete translational symmetry permits strictly quantized transport. To reset $\Delta(t)$, the soliton's center of mass must move at least one lattice period $L$, leading to the dynamical relation $x_c(T_r) - x_c(0) = q_r C\alpha = L$. The effective pumping period is $T_r = (L/ \alpha) T$, requiring $L/ \alpha$ adiabatic cycles, with an average displacement of $\alpha$ per cycle, corresponding to an $\alpha/L$ fraction of the lattice period $L$, a clear signature of fractionalized nonlinear Thouless pumping~\cite{Jurgensen2023,Tao2025}.

To gain a unified understanding of the above phenomena, we adopt a supercell approach~\cite{Jurgensen2021,Tao2025}. The nonlinear density distribution $|\psi|^2$ modifies the lattice potential, leading to a self-consistent effective Hamiltonian $H^{sc}(t)=H(t)-g|\psi(t)|^2$. For rational $\alpha$, the system can be extended into a supercell with imposed translational periodicity, allowing the Chern number of the band for $H^{sc}(t)$ to be computed. The obtained Chern number $C=1$  coincides with the topological invariant of the sliding lattice's band. For quasiperiodic cases with irrational $\alpha$, although strict quantization is broken, the nonlinear mechanism for local lattice modification remains the same, and the soliton dynamics continue to be governed by the equivalent topological structure induced by this local reconstruction; specifically, the soliton occupies a specific band and follows its Wannier center. This demonstrates that quasi-quantized pumping and fractional pumping share the same physical origin~\cite{Jurgensen2023,Tao2025}. 

Notably, the quasi-quantized pumping of solitons can also be achieved by adjusting the nonlinear strength. As shown in Figs.~\ref{fig3}(a)-(c), with moderate nonlinearity, the soliton remains in the lowest band and follows the Wannier center throughout the adiabatic cycle. To understand how nonlinearity modifies the lattice and affects the center-of-mass motion, we employ a variational approach. Using the solution of the external potential-free GP equation (\ref{GPE}), $\Psi(x)=\frac{N}{2}\operatorname{sech}[\frac{N}{2}(x-x_0)]e^{iv_0(x-x_0)}$, as a trial wave function (with $x_0$, $v_0$ the center-of-mass position and velocity), we obtain the effective equation~\cite{Fu2022,Hu2024}:
\begin{equation}\label{eff}
	\frac{d^2x_0}{dt^2}=\frac{-2\pi^3}{N}\left[\frac{4p_1\sin(4\pi x_0)}{\text{sinh}(4\pi^2/N)}+\frac{p_2\sin(2\pi x_0/\alpha-2vt)}{\alpha^2\text{sinh}(2\pi^2/(N\alpha))}\right].
\end{equation}
The equation~(\ref{eff}) indicates that the soliton's center-of-mass motion is governed by two competing amplitude factors, which depend exponentially on the norm $N$ and the lattice parameter $\alpha$, respectively. When $N$ is small, the static potential contribution is exponentially suppressed, allowing the sliding potential to dominate, leading to quasi-quantized pumping [Figs.~\ref{fig3}(a)-(c)]. When $N$ is large, the static potential dominates, localizing the soliton in the short-period lattice [Figs.~\ref{fig3}(e)(f)]. Notably, if the long lattice is fixed while the short lattice slides, the soliton pumps along the sliding direction [Figs.~\ref{fig3}(g)(h)]. Thus, by tuning $N$ and $\alpha$, one can control soliton transport~\cite{Hu2024}. Based on this, we propose two equivalent tuning pathways, provided $2\alpha>1$: (i) keep $N$ small while increasing $\alpha$; or (ii) keep $\alpha$ small while increasing $N$. Both induce nonlinear Thouless pumping, providing a clear strategy for realizing either quasi-quantized or fractional pumping of gap solitons in the lattice potential.

\emph{Conclusion}---In summary, this work systematically elucidates the nonlinear Thouless pumping mechanism for gap solitons in one-dimensional periodic and quasiperiodic superlattices. It reveals that the self-consistent potential of a soliton can reconstruct the underlying lattice, thereby providing a novel approach for the dynamic and topological manipulation of solitons. The theory is directly applicable to platforms such as ultracold atoms and photonic waveguide arrays. For instance, using the setup of Refs.~\cite{Khaykovich2002,Strecker2002}: bright solitons in a $^7\text{Li}$ cloud ($a_s \approx -1.43$ $\text{nm}$) trapped in a quasi-1D waveguide ($\omega_\perp = 2\pi \times 710$ $\text{Hz}$, $d_1 = 532$ $\text{nm}$). In our simulations, the dimensionless parameters $T = 10\pi$ and $N=10$ correspond to a time of $3.94$ $\text{ms}$ and $\tilde{N} \approx 6.69 \times 10^{3}$ atoms, respectively. Fractional Thouless pumping previously required strong nonlinearity. We show that increasing optical waveguide spacing suppresses coupling of waveguides, localizes the soliton, and allows it to reconstruct the lattice---so even weak nonlinearity induces fractional Thouless pumping (see Supplementary Material ~\cite{SM2025}). 

\begin{acknowledgments}
	\emph{Acknowledgments}---This work is supported by  he National Key Research and Development Program of China
(Grant No. 2022YFA1402704) and the programs for  NSFC of China (Grant No. 12247101).
\end{acknowledgments}   
%
\begin{widetext}
	In the Supplemental Material, the correspondence between the soliton's center of mass and the Wannier center under static and adiabatic driven conditions is examined, highlighting the differences in dynamic offset behavior between periodic and quasiperiodic systems in Section S-1, additional evidence of non-quantized soliton drift in quasiperiodic potentials through continued fraction expansions is presented, demonstrating the universality of this phenomenon and the existence of a critical approximant order in Section S-2, and the achievement of fractional topological pumping of solitons in waveguide arrays under weak nonlinearity by tuning the waveguide spacing is numerically analyzed in Section S-3.
\end{widetext}
\setcounter{equation}{0} \setcounter{figure}{0} \setcounter{table}{0} %
\renewcommand{\theequation}{S\arabic{equation}}
\renewcommand{\thefigure}{S\arabic{figure}}
\renewcommand{\bibnumfmt}[1]{[S#1]}
\renewcommand{\citenumfont}[1]{S#1}
\section{S-1.The feature of dynamical offset}
In a static lattice ($v=0$), the total potential $V_{\text{total}}(x) = V(x) + |\Psi(x)|^2$ is spatially inversion symmetric about $x_0$, i.e., $V_{\text{total}}(x_0+x) = V_{\text{total}}(x_0-x)$. It follows that both the linear lattice potential $V(x)$  and the nonlinear density distribution $|\Psi(x)|^2$ are even with respect to $x_0$. This symmetry leads to the vanishing of the integral
\begin{equation}
 x_c - x_0 = N^{-1}\int (x-x_0) |\Psi(x)|^2 dx = 0.
\end{equation}
 Consequently, the soliton's center-of-mass position $x_c\equiv N^{-1}\int x |\Psi(x)|^2 dx$ (where $N=\int |\Psi(x)|^2 dx$) coincides with the symmetry point, i.e., $x_c=x_0$. Within the single-band approximation, we adopt maximally localized Wannier functions as the basis set. For a potential $V(x)$ with symmetry center $x_0$, the density distribution $|w_m(x)|^2$ of the corresponding maximally localized Wannier function is also even about 
 $x_0$~\cite{SM_Marzari1997,SM_Marzari2012}, and its center position $X\equiv\int x|w_m(x)|^2 dx$ likewise satisfies $X=x_0$. Thus, we obtain $x_c=X=x_0$, and the dynamical offset vanishes exactly ($\Delta=x_c-X=0$). This result demonstrates that the spatial inversion symmetry of the total potential  $V_{total}(x)$ directly enforces the precise coincidence of the soliton's center of mass, the Wannier center, and the geometric symmetry point $x_0$.
  \begin{figure}[h]
 	\includegraphics[width=1\linewidth]{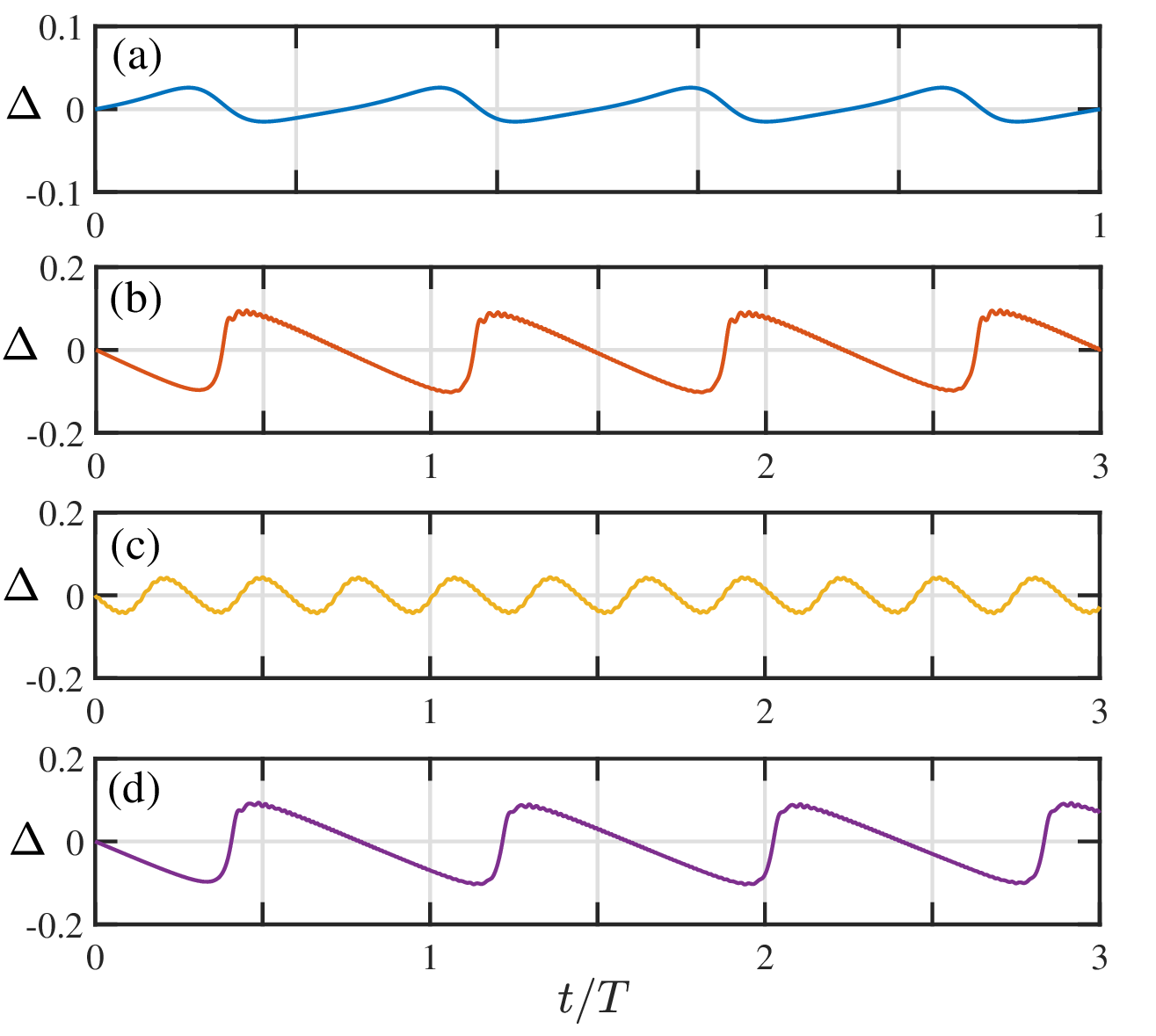}
 	\caption{We numerically computed the temporal variation of $\Delta$ under different conditions of nonlinearity strength and lattice period. (a) $N=0.2$, $\alpha=2/3$. (b) $N=7$, $\alpha=2/3$. (c) $N=0.2$, $\alpha=\sqrt{3}$. (d) $N=7$, $\alpha=\frac{\sqrt{5}-1}{2}$. The other parameters are $p_1=p_2=25$, $v=0.1$, $T=10\pi$.} 
 	\label{fig:S1}
 \end{figure}	
 
Adiabatic driving via the phase term $\varphi(t) = -vt$ breaks this spatial inversion symmetry, giving rise to a non-zero, bounded dynamical offset $\Delta(t) = x_c(t) - X(t)$. Its magnitude is determined, throughout the adiabatic process, by the instantaneous Wannier states and the corresponding soliton state distribution. Notably, the evolution of this offset differs between nonlinear periodic and nonlinear quasiperiodic systems.
 
 As shown in Figs.~\ref{fig:S1}(a) and \ref{fig:S1}(b), in a nonlinear periodic system, the discrete translational invariance of the lattice potential ensures the existence of a series of energetically equivalent, translationally invariant positions. During adiabatic evolution, the soliton's center of mass can continuously move from a symmetric position in one unit cell to an equivalent symmetric position in an adjacent cell. Consequently, after one or more complete driving cycles, the dynamical offset $\Delta$ is effectively ``reset''. This causes $\Delta$ to behave as a periodic oscillatory variable whose period is related to the adiabatic period $T$ by a specific rational ratio. In contrast, as shown in Figs.~\ref{fig:S1}(c) and \ref{fig:S1}(d), in a nonlinear quasiperiodic system, the lattice potential possesses no discrete translational symmetry, and thus no translationally invariant positions exist. As a result, the dynamical offset $\Delta$ yields a small finite net displacement per adiabatic cycle.
 
 \section{S-2 Additional Examples of Non-quantized Shifts in Quasiperiodic Potentials}
\begin{figure*}[tp]
	\includegraphics[width=1.01\linewidth]{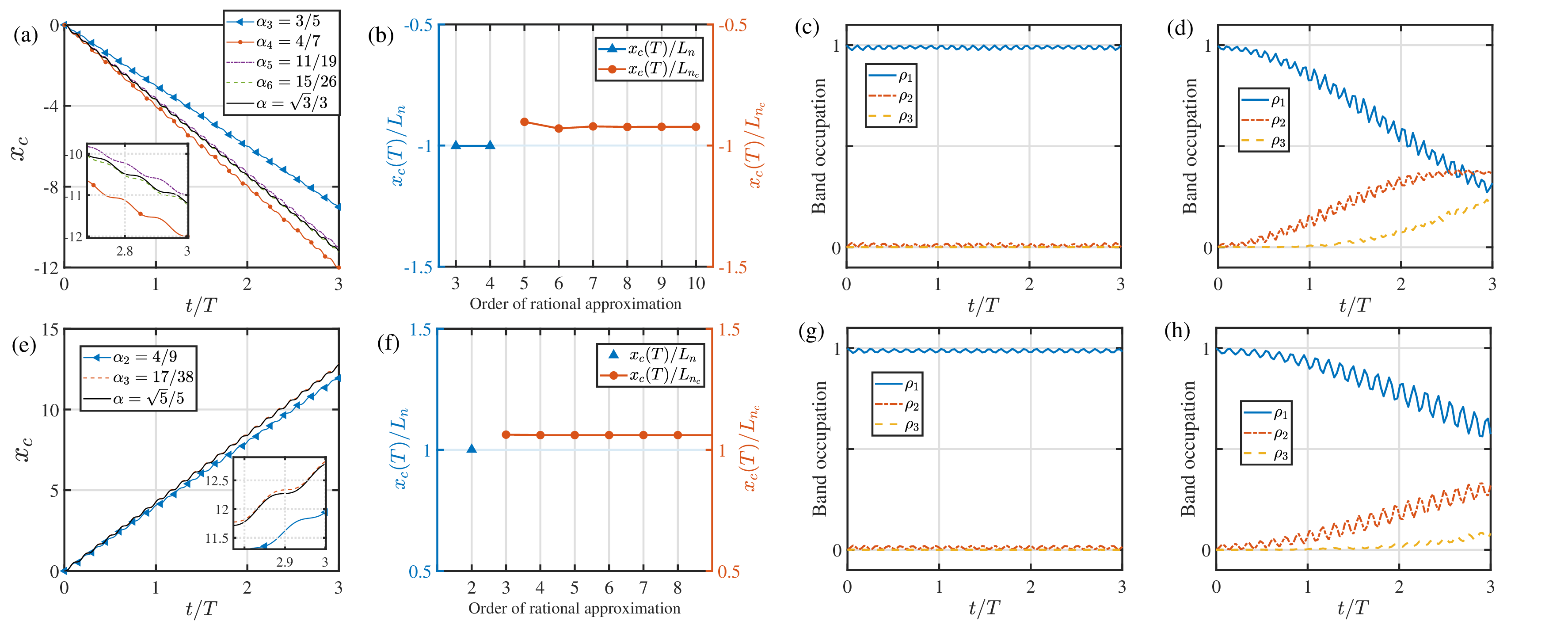}
	\centering
	\caption{Soliton dynamics under rational approximations of two irrational ratios, $\alpha=\sqrt{3}/3$ and $\alpha=\sqrt{5}/5$. The upper panels (a)-(d) correspond to $\alpha=\sqrt{3}/3$, whose continued fraction expansion is $[0;1,1,2,1,2,...]$. (a) Center-of-mass displacement of the soliton over three adiabatic cycles for successive rational approximants. (b) Identification of the critical approximant order $n_c=4$, which separates regimes of quantized transport and non-quantized drift. (c) Band occupancy of the soliton exhibiting quantized transport within the critical approximant Hamiltonian $H_{n_c}$. (d) Band occupancy of the soliton in the $n_c+1$ approximant, plotted in the band basis of $H_{n_c}$. The lower panels (e)-(h) present corresponding results for (continued fraction: $[0;2,4,4,...]$). (e) Center-of-mass displacement for its rational approximants. (f) The critical order is $n_c=2$. (g) Band occupancy under quantized transport for the $n_c$ approximant. (h) Band occupancy for the  $n_c+1$ approximant, plotted in the band basis of $H_{n_c}$. Other parameters are $p_1=p_2=25$, $N=0.2$, $v=0.1$, $T=10\pi$.}
	\label{fig:S2}
 \end{figure*}

 In the main text, we reveal that due to the absence of translational symmetry in quasiperiodic superlattices, gap solitons can exhibit non-quantized drift dynamics (e.g., as shown in Figs.~1(a)-(b)). To further validate the universality of this phenomenon, we select two representative irrational ratios, $\alpha=\sqrt{3}/3$ and $\alpha=\sqrt{5}/5$, and provide extended numerical evidence through systematic analysis of soliton dynamics in their sequences of rational approximants. All results are summarized in Figs.~\ref{fig:S2}, where Figs.~\ref{fig:S2}(a)-(d) correspond to $\alpha=\sqrt{3}/3$, and Figs.~\ref{fig:S2}(e)-(h) correspond to $\alpha=\sqrt{5}/5$.
 
Figures~\ref{fig:S2}(a) and \ref{fig:S2}(e) display the change in the soliton's center-of-mass displacement over three adiabatic cycles as a function of the approximant order $n$ for two different parameter sets, respectively. Both panels clearly reveal the existence of a critical order $n_c$, which separates the dynamical behavior into two distinct regimes. For approximant orders $n\leq n_c$, the displacement exhibits the characteristics of quantized transport. Once the order exceeds $n_c$ (i.e., $n>n_c$), the displacement deviates from quantized values and, with increasing $n$, converges to a common, non-quantized value. This convergence indicates that in the limit approaching the quasiperiodic potential ($n\to \infty$), the dynamical outcomes from different high-order approximants become consistent, collectively pointing to a non-quantized drift value. This demonstrates that non-quantized drift is a universal phenomenon emerging in quasiperiodic systems under different parameters.

To characterize this transition more precisely and to highlight its connection to topology, Figs.~\ref{fig:S2}(b) and \ref{fig:S2}(f) employ dual vertical axes for quantitative comparison. For the regime $n\leq n_c$ (left axis), we scale the displacement $x_c(T)$ by the superlattice period $L_n$ of the corresponding approximant, obtaining the ratio $x_c(T)/L_n$. The results show that for $\alpha=\sqrt{3}/3$, this ratio remains locked near $-1$ [Fig.~\ref{fig:S2}(b)], whereas for  $\alpha=\sqrt{5}/5$, it stays locked near $1$ [Fig.~\ref{fig:S2}(f)]. According to the theory of nonlinear Thouless pumping, this ratio directly corresponds to the Chern number $C$ of the band predominantly occupied by the soliton, i.e., $x_c(T)=CL_n$ (with $x_c(0)=0$). Consequently, the results confirm that for systems below the critical order ($n\leq n_c$), the Chern numbers are $C=-1$ for the $\alpha=\sqrt{3}/3$ rational approximants and $C=1$ for the $\alpha=\sqrt{5}/5$ rational approximants. This directly verifies the quantized nature and topological origin of the transport in this regime.

For the regime $n> n_c$ (right axis), we instead scale the displacement by the period $L_{n_c}$ of the critical approximant system, plotting the ratio $x_c(T)/L_{n_c}$. As shown in Figs.~\ref{fig:S2}(b) and \ref{fig:S2}(f), this ratio converges to a non-integer value. This indicates that beyond the critical order, the soliton net displacement per adiabatic cycle, while no longer quantized, remains on the order of the lattice period $L_{n_c}$ of the critical Hamiltonian $H_{n_c}$. It is noteworthy that the net drift direction still exhibits a clear regularity. By analyzing two representative systems with opposite Chern numbers $C = \pm 1$ [Figs.~\ref{fig:S2}(b) and \ref{fig:S2}(f)], we find that the transport direction aligns with the direction dictated by the Chern number of their respective critical bands, with the direction set by the sign of the Chern number and the fixed driving protocol.

This transition from quantized to non-quantized behavior can be traced to the establishment and breakdown of the adiabatic connection between the soliton and the specific topologically nontrivial band in the critical approximant $H_{n_c}$. For low-order approximants $n\leq n_c$, the soliton stably occupies the lowest band of the corresponding rational approximant during the adiabatic cycle. Consequently, the soliton's adiabatic transport responds to the band topology, with its displacement governed by the corresponding Chern number, thereby leading to quantized pumping [see Fig.~\ref{fig:S2}(c) and \ref{fig:S2}(g)]. In contrast, for a supercritical approximant ($n>n_c$), the occupancy of the same lowest band decays and transfers over time [see Figs.~\ref{fig:S2}(d) and \ref{fig:S2}(h)]. This confirms that the higher-order perturbation potential $W^m_{n_c}=H_m-H_n$ breaks the adiabatic connection, thereby driving the system into a regime of non-quantized drift. 

\section{S-3 Fractional Thouless pumping of solitons under weak nonlinearity in photonic waveguide arrays}
As indicated in the main text, adjusting the nonlinear strength or the lattice period can induce quasi-quantized or fractional Thouless pumping of gap solitons (see Fig.~3 and the related discussion). To verify the universality of this mechanism, this section further examines a representative photonic platform--coupled optical waveguide arrays--which is widely employed in topological photonics for studying nonlinear transport phenomena. Previous work has shown that this system can support fractional Thouless pumping of solitons under strong nonlinearity. Through systematic numerical simulations, this Supplemental Material shows that even under weak nonlinearity, fractional Thouless pumping of solitons can still be achieved through a geometric design.
\begin{figure}[h]
	\includegraphics[width=1.01\linewidth]{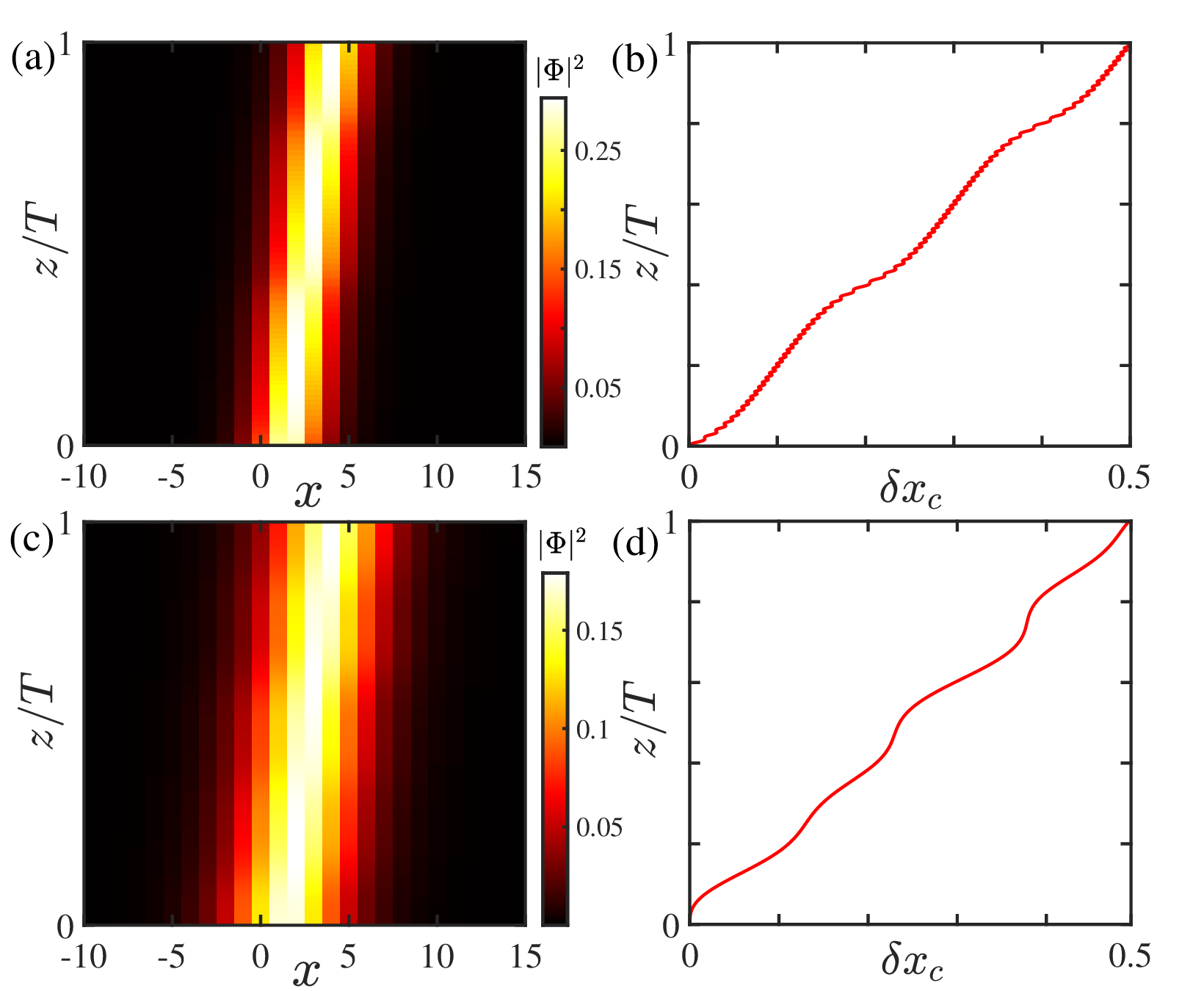}
	\centering
	\caption{Two pathways to nonlinear fractional Thouless pumping. The nearest-neighbor hoppings are periodically modulated in $z$ with period $T=2\pi/\Omega$ according to $J_n(z)=J+K\cdot\cos\left(2\pi q(n-1)/p+\Omega z\right)$. In all simulations, the parameters are fixed as $q=2$, $p=5$, $K=0.01$, and $\Omega=0.01$. Panels (a) and (b) correspond to strong nonlinearity ($gN=2.1$) with $J=1$. Over one adiabatic cycle, the soliton traverses half a unit cell. Panels (c) and (d) correspond to weak nonlinearity ($gN=0.2$) with reduced coupling $J=0.15$. The smaller $J$ lowers the required nonlinearity strength, yet the soliton still completes half a unit cell per cycle. The soliton displacement is quantified by the normalized center-of-mass shift $\delta x_c(z)=[x_c(z)-x_c(0)]/p$, with $x_c(z)=\sum n|\phi_n(z)|^2/N$, where $p$ is the unit-cell length.}
	\label{fig:S3}
\end{figure}	

For a one-dimensional array of coupled optical waveguides, under the paraxial approximation, the propagation of light is described by the following discrete nonlinear Schr\"{o}dinger equation~\cite{SM_Jurgensen2023,SM_Jurgensen2021}:
\begin{equation}
\begin{aligned}
i\frac{\partial}{\partial z}\phi_n(z)=&-J_n(z)\phi_{n+1}(z)-J_{n-1}(z)\phi_{n-1}(z)\\
&-g|\phi_n(z)|^2\phi_n(z),
\label{DNLS}
\end{aligned}
\end{equation}
where $\phi_n(z)$ is the electric field envelope at propagation distance 
$z$ for waveguide $n$, and $g>0$ represents the strength of the focusing Kerr nonlinearity. The linear part of Eq.~(\ref{DNLS}) is described by an off-diagonal Aubry-Andr\'{e}-Harper (AAH) mode, with the nearest-neighbor coupling strengths $J_n(z)$ modulated periodically in $z$. This modulation realizes a two-dimensional Chern insulator in the synthetic dimension $(n,z)$, enabling topological transport. Experimentally, the periodic profile of $J_n(z)$ is achieved by carefully designing and periodically varying the transverse separation $d(z)$ between adjacent waveguides along $z$. The coupling strength $J$ depends sensitively and exponentially on the instantaneous center-to-center distance $d$, with an empirical fit given by $J(d)/\mathrm{mm}^{-1}=6.672\exp(-234.8\mathrm{~}d/\mathrm{mm})$~\cite{SM_Jurgensen2023}. This relation allows the effective coupling strength $J_n(z)$ to be tuned continuously and precisely over a wide range through slight, periodic adjustments of the waveguide separation $d$, thereby providing a direct and flexible experimental means to realize and manipulate topological phases in synthetic dimensions.

In nonlinear systems, when the localization width of a soliton is much smaller than the lattice period, the soliton becomes tightly confined to only a few lattice sites, and its dynamical behavior is governed by this highly localized region rather than by the global periodicity of the lattice. Within this region, the local potential variation induced by the soliton solution modifies the linear Hamiltonian, introducing a nontrivial topological structure~\cite{SM_Tao2025}, which in turn causes the soliton to exhibit fractional Thouless pumping during adiabatic evolution~\cite{SM_Jurgensen2023}. In photonic waveguide arrays, there are two primary physical pathways to achieve fractional Thouless pumping:
\begin{enumerate}
	\item \textbf{Enhancing nonlinearity to induce strong self-localization}. This approach has been extensively studied both theoretically and experimentally~\cite{SM_Jurgensen2023} (as shown in Figs.~\ref{fig:S3}(a),(b), corresponding to the strongly nonlinear regime).
	\item \textbf{Exponentially suppressing the coupling strength $J$ by increasing the waveguide separation}. As illustrated in Figs.~\ref{fig:S3}(c),(d), when $J$ is reduced to 0.15, solitons still exhibit clear fractional pumping behavior even under a weak nonlinear strength of only $gN=0.2$. This is because suppressing $J$ strongly inhibits the linear diffraction of the optical field, facilitating the formation of localized states with a width much smaller than the lattice period. Consequently, fractionalized transport can be achieved even under weak nonlinearity.
\end{enumerate}
It is worth noting that the latter pathway reveals the compensatory role of geometric design over nonlinear requirements: by tailoring the waveguide spacing to weaken linear coupling, the dependence on the material's nonlinear coefficient for achieving fractional pumping can be significantly reduced. This substantially expands the potential for realizing such topological effects in photonic platforms with inherently low nonlinearities.

The above results are consistent with the conclusions presented in the main text: The modification of the underlying lattice by the soliton's self-consistent potential is key to inducing topological pumping, and it can be equivalently achieved by tuning system parameters, the coupling strength $J$ or the nonlinear coefficient $gN$. This further confirms the universality of the mechanism proposed in the main text across different physical platforms, namely ultracold atoms and photonic waveguides, and provides a feasible route toward achieving topological control in systems with weak nonlinearity.

\end{document}